# Multistep Frequency Response Optimized Integrators and Their Application to Accelerating a Power System Transient Simulation Scheme

Sheng Lei[1,2], Student Member, IEEE, and Alexander Flueck[1], Senior Member, IEEE

[1]Department of Electrical and Computer Engineering  
Illinois Institute of Technology  
Chicago, IL, USA

[2]Mathematics and Computer Science Division  
Argonne National Laboratory  
Lemont, IL, USA

Email: slei3@hawk.iit.edu, and flueck@iit.edu

*Abstract*—**This paper proposes several explicit and implicit multistep frequency response optimized integrators considering first or second order derivative. A prediction-based method aiming at accelerating a novel power system transient simulation scheme without impacting its accuracy is further put forward utilizing the proposed numerical integrators and some others available in the literature. Case studies verify the effectiveness of the proposed prediction method. Although they are utilized to accelerate the simulation scheme in this paper, the proposed numerical integrators are in fact general-purpose and can be applied to other areas.**

## I. Introduction

Stability analysis of general unbalanced power systems is attracting increasing research interest and effort, where electromagnetic transient (EMT) simulation has been used [1]-[2] due to its ability to model unbalanced systems in detail and simulate comprehensive time domain dynamics [3]-[4]. Nevertheless traditional EMT simulation is unnecessarily inefficient for power system stability studies. Small step sizes have to be adopted due to the presence of the utility frequency (50 or 60 Hz) in the waveforms. Otherwise the commonly used implicit trapezoidal method will introduce significant error or even lead to divergence if large step sizes are directly enforced [5]-[8].

Some schemes have been proposed in the literature to enable large step sizes in EMT-like simulation. Shifted frequency analysis (SFA) [7] computes the frequency shifted analytic signals instead of the natural signals themselves. The frequency shifted analytic signals are slow variants so that the conventional implicit trapezoidal method can be applied with large step sizes. Frequency-adaptive simulation of transients (FAST) [8] switches between the analytic signals and the frequency shifted ones in an adaptive manner. In [9], the authors put forward a novel transient simulation scheme based on frequency response optimized integrators considering second order derivative. The frequency response of the numerical integrators is optimized according to the frequency spectrum of individual variables so that large step sizes are enabled without scarifying accuracy. Detailed analysis on properties of the numerical integrators in [9], including error and numerical stability, can be found in [10].

Several techniques may be applied to accelerate transient simulation. Dynamic equivalencing [11] may be performed on the portions of the studied system which are not of concern. In EMT-transient stability (TS) hybrid simulation [12], part of the studied system is simulated with computationally less expensive TS simulation. The studied system may be divided into fast subsystems and slow ones and simulated with different step sizes respectively [13]. A latency of one time step may be inserted into the solution process at each time step so that the studied system is decoupled into subsystems which can be solved separately [3]-[4]. Another category of techniques is parallel computing [14].

It is of interest to further accelerate the novel transient simulation scheme [9] without impacting its accuracy. This paper aims at achieving the acceleration by predicting the variables at each time step with multistep frequency response optimized integrators, which are to be corrected by later calculation. Contributions of this paper are threefold. First, several multistep frequency response optimized integrators are proposed. Second, a method to predict the variables at each time step is put forward to accelerate transient simulation based on the proposed numerical integrators and some others available in the literature. Third, the efficiency of the novel transient simulation scheme is enhanced by the proposed prediction method while its accuracy remains intact.

The rest of this paper is organized as follows. Section II introduces explicit two-step numerical integrators considering second order derivative. Explicit and implicit multistep numerical integrators considering first order derivative are introduced in Sections III and IV respectively. Section V introduces methods to provide an initial guess for Newton's method. Section VI verifies the proposed prediction method via case studies. Section VII concludes the paper and points out some directions for future research.

## II. EXPLICIT TWO-STEP FREQUENCY RESPONSE OPTIMIZED INTEGRATORS CONSIDERING SECOND ORDER DERIVATIVE

Consider an ordinary differential equation (ODE)

$$\dot{x} = f(t,x,u) \quad (1)$$

where $x$ is the state variable; $t$ is the time instant; $u$ is the input; $f$ is a function of $t$, $x$ and $u$. An explicit two-step numerical integrator considering second order derivative may be applied to discretize (1) so that (1) can be solved numerically

$$x_t = x_{t-h} + b_{-1}\dot{x}_{t-h} + b_{-2}\dot{x}_{t-2h} + c_{-1}\ddot{x}_{t-h} + c_{-2}\ddot{x}_{t-2h} \quad (2)$$

where $h$ is the step size; $b_{-1}$, $b_{-2}$, $c_{-1}$ and $c_{-2}$ are coefficients. A specific selection for these coefficients defines a numerical integrator. The second order derivative of $x$ in (2) can be calculated by taking derivative on both sides of (1) as

$$\ddot{x} = \frac{\partial f}{\partial t} + \frac{\partial f}{\partial x}\dot{x} + \frac{\partial f}{\partial u}\dot{u} \quad (3)$$

Obviously (3) requires the derivative of $u$. Such information may be externally provided or numerically calculated given $u$.

Reference [5] proposed selecting the coefficients of a single-step numerical integrator considering first order derivative according to desired frequency response of the error. That idea is generalized in this paper to select the coefficients in (2). Performing Laplace transform on both sides of (2)

$$X = Xe^{-sh} + b_{-1}sXe^{-sh} + b_{-2}sXe^{-2sh} + c_{-1}s^2Xe^{-sh} + c_{-2}s^2Xe^{-2sh} \quad (4)$$

The $s$-domain error is

$$X - (Xe^{-sh} + b_{-1}sXe^{-sh} + b_{-2}sXe^{-2sh} + c_{-1}s^2Xe^{-sh} + c_{-2}s^2Xe^{-2sh}) \quad (5)$$

The $s$-domain relative error is

$$1 - e^{-sh} - b_{-1}se^{-sh} - b_{-2}se^{-2sh} - c_{-1}s^2e^{-sh} - c_{-2}s^2e^{-2sh} \quad (6)$$

If the coefficients are selected so that $j\omega_{select}$ and $-j\omega_{select}$ are both a root of the error expression (6), the resulting numerical integrator introduces no error at the nonzero angular frequency $\omega_{select}$ despite the specific value of the step size. If 0 is a root of (6), the numerical integrator is accurate for slow variants. Higher multiplicity of a root implies higher accuracy at the corresponding frequency. A numerical integrator with desired error frequency response can be designed by constructing and solving an equation set regarding the coefficients considering the root conditions. An example will be presented later in this section.

The state variable of some ODEs may have a nonzero dominant frequency component, such as the utility frequency in power system voltage and current waveforms. Accordingly $\omega_{select}$ may be specified at the dominant frequency to reduce the overall error introduced by the discretization of ODEs. The frequency response of the numerical integrator is thus optimized from the numerical error viewpoint. Consequently large step sizes are enabled to increase efficiency without sacrificing accuracy. Similarly numerical integrators having 0 as a root of the error expression (6) can also be understood as frequency response optimized for slow variants.

### A. Modified Generalized Two-Step Adams-Bashforth Method

This paper proposes the modified generalized two-step Adams-Bashforth method, for which $j\omega_{select}$ and $-j\omega_{select}$ are made a single root while 0 is made a triple root of the error expression (6). Hence the numerical integrator is accurate for variables with a dominant frequency component at $\omega_{select}$. It is also rather accurate for slow variants.

Specifically the following equation set is constructed

$$\begin{aligned}
&(1-e^{-sh}-b_{-1}se^{-sh}-c_{-1}s^2e^{-sh}-b_{-2}se^{-2sh}-c_{-2}s^2e^{-2sh})\big|_{s=-j\omega_{select}}=0\\
&(1-e^{-sh}-b_{-1}se^{-sh}-c_{-1}s^2e^{-sh}-b_{-2}se^{-2sh}-c_{-2}s^2e^{-2sh})\big|_{s=j\omega_{select}}=0\\
&(1-e^{-sh}-b_{-1}se^{-sh}-c_{-1}s^2e^{-sh}-b_{-2}se^{-2sh}-c_{-2}s^2e^{-2sh})\big|_{s=0}=0\\
&(\frac{d}{ds}(1-e^{-sh}-b_{-1}se^{-sh}-c_{-1}s^2e^{-sh}-b_{-2}se^{-2sh}-c_{-2}s^2e^{-2sh}))\big|_{s=0}=0\\
&(\frac{d^2}{ds^2}(1-e^{-sh}-b_{-1}se^{-sh}-c_{-1}s^2e^{-sh}-b_{-2}se^{-2sh}-c_{-2}s^2e^{-2sh}))\big|_{s=0}=0
\end{aligned} \quad (7)$$

The solution is

$$b_{-1} = \frac{num_{b_{-1}}}{den_b}, \quad b_{-2} = \frac{num_{b_{-2}}}{den_b}, \quad c_{-1} = \frac{num_{c_{-1}}}{den_c}, \quad c_{-2} = \frac{num_{c_{-2}}}{den_c} \quad (8)$$

where

$$\begin{aligned}
num_{b_{-1}} &= 12h^3\omega_{select}^4 + 32\omega_{select}\sin(\omega_{select}h) - 32h^2\omega_{select}^3\sin(\omega_{select}h)\\
&-32\omega_{select}\cos(\omega_{select}h)\sin(\omega_{select}h) - 16h\omega_{select}^2\cos(\omega_{select}h)\\
&+12h^3\omega_{select}^4\cos(\omega_{select}h) + 16h\omega_{select}^2\cos^2(\omega_{select}h)
\end{aligned} \quad (9)$$

$$\begin{aligned}
num_{b_{-2}} &= 32h\omega_{select}^2 - 4h^3\omega_{select}^4 - 32\omega_{select}\sin(\omega_{select}h)\\
&+32\omega_{select}\cos(\omega_{select}h)\sin(\omega_{select}h) - 16h\omega_{select}^2\cos(\omega_{select}h)\\
&-4h^3\omega_{select}^4\cos(\omega_{select}h) - 16h\omega_{select}^2\cos^2(\omega_{select}h)
\end{aligned} \quad (10)$$

$$\begin{aligned}
den_b &= 4\omega_{select}^2(-2\sin(\omega_{select}h) + \omega_{select}h\cos(\omega_{select}h) + \omega_{select}h)^2\\
&+4\omega_{select}^2(\omega_{select}h\sin(\omega_{select}h) + 2\cos(\omega_{select}h) - 2)^2
\end{aligned} \quad (11)$$

$$\begin{aligned}
num_{c_{-1}} &= \omega_{select}^2(16(\omega_{select}h)^2 - 8(\omega_{select}h)^3\sin(\omega_{select}h)\\
&+96\omega_{select}h\cos(\omega_{select}h)\sin(\omega_{select}h)\\
&+24(\omega_{select}h)^3\cos(\omega_{select}h)\sin(\omega_{select}h)\\
&-96\omega_{select}h\cos^2(\omega_{select}h)\sin(\omega_{select}h)\\
&-64\cos(\omega_{select}h) - 96(\omega_{select}h)^2\cos(\omega_{select}h)\\
&+128\cos^2(\omega_{select}h) + 48(\omega_{select}h)^2\cos^2(\omega_{select}h)\\
&-64\cos^3(\omega_{select}h) + 32(\omega_{select}h)^2\cos^3(\omega_{select}h))
\end{aligned} \quad (12)$$

$$\begin{aligned}
num_{c_{-2}} &= -8\omega_{select}^2(-10(\omega_{select}h)^2\\
&+4\omega_{select}h\sin(\omega_{select}h) + 3(\omega_{select}h)^3\sin(\omega_{select}h)\\
&-2\omega_{select}h\sin(2\omega_{select}h) - \frac{1}{2}(\omega_{select}h)^3\sin(2\omega_{select}h)\\
&-8\cos(\omega_{select}h) + 8(\omega_{select}h)^2\cos(\omega_{select}h)\\
&+16\cos^2(\omega_{select}h) + 2(\omega_{select}h)^2\cos^2(\omega_{select}h)\\
&+4\omega_{select}h\sin^3(\omega_{select}h) - 8\cos^3(\omega_{select}h))
\end{aligned} \quad (13)$$

$$\begin{aligned}
den_c &= -16\omega_{select}^4(\cos(\omega_{select}h) - 1)(-4\omega_{select}h\sin(\omega_{select}h)\\
&-4\cos(\omega_{select}h) + (\omega_{select}h)^2\cos(\omega_{select}h) + 4 + (\omega_{select}h)^2)
\end{aligned} \quad (14)$$

### B. Classical Generalized Two-Step Adams-Bashforth Method

A numerical integrator has the coefficients [15]

$$b_{-1} = -\frac{1}{2}h, \ b_{-2} = \frac{3}{2}h, \ c_{-1} = \frac{17}{12}h^2, \ c_{-2} = \frac{7}{12}h^2 \qquad (15)$$

This numerical integrator is referred to as the classical generalized two-step Adams-Bashforth method in this paper. It can be verified that 0 is a quintuple root of the error expression (6). Therefore the numerical integrator is highly accurate for slow variants.

## III. FREQUENCY RESPONSE OPTIMIZED ADAMS-BASHFORTH METHODS

Applying a three-step Adams-Bashforth-like numerical integrator [15] to (1), the discretized equation is

$$x_t = a_{-1}x_{t-h} + b_{-1}\dot{x}_{t-h} + b_{-2}\dot{x}_{t-2h} + b_{-3}\dot{x}_{t-3h} \qquad (16)$$

The coefficients in (16) can also be selected so that the error frequency response of the resulting numerical integrator is optimized for variables with a nonzero dominant frequency component or slow variants. The desired selection is obtained via the same method as that detailed in Section II.

Three-step frequency response optimized Adams-Bashforth methods are introduced as follows in this section, which are adopted by the proposed prediction method. The two-step versions are given in Appendix A for completeness.

### A. Modified Three-Step Adams-Bashforth Method

This paper proposes the modified three-step Adams-Bashforth method of which the coefficients are

$$a_{-1} = 1$$

$$b_{-1} = \frac{-2\cos(2\omega_{select}h) + \omega_{select}h\cot(\frac{\omega_{select}h}{2})}{2\omega_{select}\sin(\omega_{select}h)}$$

$$b_{-2} = \frac{4}{\omega_{select}}\cos^2(\frac{\omega_{select}h}{2})\cot(\frac{\omega_{select}h}{2}) \qquad (17)$$

$$-\frac{3}{\omega_{select}}\cot(\frac{\omega_{select}h}{2}) - h\frac{\cos(\omega_{select}h)}{1-\cos(\omega_{select}h)}$$

$$b_{-3} = \frac{-2\cos(\omega_{select}h) + \omega_{select}h\cot(\frac{\omega_{select}h}{2})}{2\omega_{select}\sin(\omega_{select}h)}$$

With these coefficients, $j\omega_{select}$ and $-j\omega_{select}$ are made a single root while 0 is made a double root of the error expression (6). Therefore this numerical integrator is accurate for variables with a dominant frequency component at $\omega_{select}$ and slow variants.

### B. Classical Three-Step Adams-Bashforth Method

The classical three-step Adams-Bashforth method is expressed as [15], [16]

$$a_{-1} = 1, \ b_{-1} = \frac{23}{12}h, \ b_{-2} = -\frac{4}{3}h, \ b_{-3} = \frac{5}{12}h \qquad (18)$$

It can be verified that 0 is a quadruple root of the error expression (6). Therefore this numerical integrator is highly accurate for slow variants.

## IV. FREQUENCY RESPONSE OPTIMIZED BACKWARD DIFFERENTIATION FORMULAS

The ODE (1) is discretized by a three-step backward differentiation formula-like numerical integrator [15] as

$$x_t = a_{-1}x_{t-h} + a_{-2}x_{t-2h} + a_{-3}x_{t-3h} + b_0\dot{x}_t \qquad (19)$$

Again the error frequency response of the numerical integrator can be optimized for variables with a nonzero dominant frequency component or slow variants by properly selecting the coefficients. The method to select these coefficients so as to meet the requirements is detailed in Section II.

Three-step frequency response optimized backward differentiation formulas adopted by the proposed prediction method are introduced as follows in this section. For completeness, the two-step versions are given in Appendix B.

### A. Modified Three-Step Backward Differentiation Formula

This paper proposes the modified three-step backward differentiation formula which has the following coefficients

$$a_{-1} = \frac{num_{a_{-1}}}{4(-\omega_{select}h + \sin(\omega_{select}h) + 2\omega_{select}h\sin^2(\omega_{select}h))^2}$$

$$a_{-2} = \frac{num_{a_{-2}}}{4(-\omega_{select}h + \sin(\omega_{select}h) + 2\omega_{select}h\sin^2(\omega_{select}h))^2}$$

$$a_{-3} = \frac{num_{a_{-3}}}{4(-\omega_{select}h + \sin(\omega_{select}h) + 2\omega_{select}h\sin^2(\omega_{select}h))^2} \qquad (20)$$

$$b_0 = \frac{num_{b_0}}{4(-\omega_{select}h + \sin(\omega_{select}h) + 2\omega_{select}h\sin^2(\omega_{select}h))^2}$$

where

$$num_{a_{-1}} = 4(\omega_{select}h)^2 + 2 + 2\cos(\omega_{select}h)$$
$$+ 2(\omega_{select}h)^2\cos(\omega_{select}h) - 2\cos(2\omega_{select}h) - 2\cos(3\omega_{select}h)$$
$$+ 2(\omega_{select}h)^2\cos(3\omega_{select}h) + 4(\omega_{select}h)^2\cos(4\omega_{select}h) \qquad (21)$$
$$+ 6\omega_{select}h\sin(\omega_{select}h) - 2\omega_{select}h\sin(2\omega_{select}h)$$
$$- 6\omega_{select}h\sin(3\omega_{select}h) - 2\omega_{select}h\sin(4\omega_{select}h)$$

$$num_{a_{-2}} = -4(\omega_{select}h)^2 - 4$$
$$- 8\cos(\omega_{select}h) + 8(\omega_{select}h)^2\cos(\omega_{select}h)$$
$$+ 4\cos^2(\omega_{select}h) + 16(\omega_{select}h)^2\cos^2(\omega_{select}h)$$
$$+ 8\cos^3(\omega_{select}h) - 16(\omega_{select}h)^2\cos^3(\omega_{select}h) \qquad (22)$$
$$- 16(\omega_{select}h)^2\cos^4(\omega_{select}h) - 8(\omega_{select}h)\sin(\omega_{select}h)$$
$$+ 16\omega_{select}h\cos^2(\omega_{select}h)\sin(\omega_{select}h)$$
$$+ 16\omega_{select}h\cos^3(\omega_{select}h)\sin(\omega_{select}h)$$

$$num_{a_{-3}} = 4 - 4(\omega_{select}h)^2\cos(\omega_{select}h) - 4\cos^2(\omega_{select}h)$$
$$+ 8(\omega_{select}h)^2\cos^3(\omega_{select}h) + 4\omega_{select}h\sin(\omega_{select}h) \qquad (23)$$
$$- 4\omega_{select}h\cos(\omega_{select}h)\sin(\omega_{select}h)$$
$$- 8\omega_{select}h\cos^2(\omega_{select}h)\sin(\omega_{select}h)$$

$$num_{b_0} = 2h(2 - \cos(\omega_{select}h) - 2\cos(2\omega_{select}h) + \cos(3\omega_{select}h)$$
$$+ 2\omega_{select}h\sin(\omega_{select}h) - 2\omega_{select}h\sin(3\omega_{select}h) \qquad (24)$$
$$+ \omega_{select}h\sin(4\omega_{select}h))$$

$j\omega_{select}$ and $-j\omega_{select}$ are made a single root while 0 is made a double root of the error expression (6) with these coefficients. Therefore this numerical integrator is accurate for variables dominated by the $\omega_{select}$ component and slow variants.

### B. Classical Three-Step Backward Differentiation Formula

The classical three-step backward differentiation formula is expressed as [15], [16]

$$a_{-1} = \frac{18}{11}, \ a_{-2} = -\frac{9}{11}, \ a_{-3} = \frac{2}{11}, \ b_0 = \frac{6}{11}h \qquad (25)$$

It can be verified that 0 is a quadruple root of the error expression (6). Therefore this numerical integrator is highly accurate for slow variants.

## V. METHODS FOR INITIAL GUESS

### A. Review of the Novel Transient Simulation Scheme

The novel transient simulation scheme [9] adopts a fixed step size during a simulation run, which is the same as commonly used in EMT simulation. It solves the whole system simultaneously at each time step with Newton's method, ensuring consistency to achieve high fidelity. Discontinuities are properly dealt with by a similar technique to the Critical Damping Adjustment (CDA) [17] where two time steps immediately after a discontinuity are calculated with half the user-specified step size and different numerical integrators. Consequently time steps are classified as normal ones and half ones.

When utilizing numerical integrators considering second order derivative, the derivative of the input to an ODE is required [9], similar to (3) in this paper. The input may be an algebraic state variable of the studied system. In this case, the derivative of that variable has to be calculated at each time step, as well as the variable itself.

### B. Methods for Initial Guess for Newton's Method

To start Newton's method, an initial guess of the unknowns is necessary [16]. A naïve method to provide the initial guess is to use the values at the previous time step directly. On the other hand, if the initial guess is close enough to the final solution, the convergence will be sped up [16], hopefully leading to acceleration of the overall simulation. Based on such observation, this paper proposes a method to provide the initial guess by predicting the values at the current time step with the available values at previous time steps.

Specifically, if the current time step and the consecutive two previous ones are all normal, the proposed prediction method works as follows; otherwise the values at the previous time step are used. Before the first Newton iteration, the value of state variables is predicted.

The differential state variables are predicted by an explicit two-step frequency response optimized integrator considering second order derivative. The algebraic state variables are predicted by a frequency response optimized Adams-Bashforth method; then they are used in the prediction of the derivative. The derivative is predicted by using a frequency response optimized backward differentiation formula as a differentiator. According to (19)

$$\dot{x}_t = \frac{1}{b_0}(x_t - a_{-1}x_{t-h} - a_{-2}x_{t-2h} - a_{-3}x_{t-3h}) \quad (26)$$

Selection of numerical integrators is achieved in a case-by-case manner according to the frequency spectrum of individual variables. Details are given in Sections II-IV. It should be pointed out that some rotor quantities, such as the rotor flux linkages, may contain significant second order harmonics if the system is unbalanced [18]. For these variables, a frequency response optimized integrator should be adopted with $\omega_{select}$ specified at double the utility frequency. Although the formulas for the coefficients look complicated as shown in Sections II-IV, they are calculated only once before entering the time loop; then the stored values are used.

## VI. CASE STUDIES

The proposed prediction method is added as an optional functionality to the MATLAB implementation of the novel transient simulation scheme [9]. Its effectiveness is to be verified via case studies in this section. The same test system as that in [9] is adopted in this paper, which is a modified version of the well-known 3-machine 9-bus system [19]. Bus parameters, branch parameters and power flow data can be found in [19]. Generator dynamic parameters can be found in [9]. They are not provided in this paper due to space limitation. Unbalance is introduced into the system by non-uniform allocation of loads on individual phases as detailed in [9]. At 0.1 s, a Phase-B-and-Phase-C-to-ground fault is applied at Bus 9 with a fault resistance of 0.001 p.u.. At 0.3 s, the fault is cleared. Duration of the simulation runs is from 0.0 s to 5.0 s. The novel transient simulation scheme has already been validated in [9] with this test system.

The novel transient simulation scheme executes simulation runs using the proposed prediction method or the naïve method for initial guess as discussed in Section V. Average number of iterations per time step (ANITS) is calculated for each simulation run. ANITS results from the two methods given different step sizes and user-specified tolerances for Newton's method regarding the infinite norm of the residual [16] are listed in Table I. Time consumption results from the two methods given different conditions are listed in Table II.

TABLE I. ANITS COMPARISON OF THE TWO METHODS

| Step Size (μs) | Tolerance (p.u.) | | | |
|---|---|---|---|---|
| | $10^{-8}$ | | $10^{-6}$ | |
| | Proposed | Naïve | Proposed | Naïve |
| 125 | 1.01 | 2.00 | 1.00 | 2.00 |
| 250 | 1.01 | 2.00 | 1.00 | 2.00 |
| 500 | 1.05 | 2.45 | 1.00 | 2.00 |
| 1000 | 1.82 | 2.98 | 1.05 | 2.16 |
| 2000 | 2.00 | 2.98 | 1.89 | 2.97 |
| 4000 | 2.05 | 3.00 | 2.01 | 2.98 |

TABLE II. TIME CONSUMPTION COMPARISON OF THE TWO METHODS

| Step Size (μs) | Tolerance (p.u.) | | | |
|---|---|---|---|---|
| | $10^{-8}$ | | $10^{-6}$ | |
| | Proposed | Naïve | Proposed | Naïve |
| 125 | 222.84 | 264.75 | 222.97 | 265.20 |
| 250 | 112.25 | 132.77 | 112.05 | 132.61 |
| 500 | 55.35 | 72.10 | 55.16 | 65.63 |
| 1000 | 36.53 | 41.25 | 28.33 | 33.79 |
| 2000 | 19.23 | 20.71 | 17.92 | 20.66 |
| 4000 | 9.56 | 10.73 | 9.44 | 10.70 |

As can be seen from Table I, the proposed prediction method clearly speeds up the convergence of Newton's method in that its ANITS is always smaller than that of the naïve method given the same condition. Although the prediction method adds more computation compared to the naïve method, the additional cost pays off with acceleration of the overall simulation and thus enhancement in computational efficiency. As can be seen from Table II, the proposed prediction method consumes less time than the naïve method under the same condition. Therefore the effectiveness of the proposed prediction method is justified.

## VII. Conclusion and Future Work

Several multistep frequency response optimized integrators are proposed in this paper. The proposed numerical integrators and some others are applied to predict the variables so as to accelerate the novel transient simulation scheme [9]. Case studies verify the prediction method.

In the future, more delicate methods may be investigated to further enhance the efficiency of the novel transient simulation scheme. The proposed numerical integrators are in fact general-purpose. Their numerical properties and applications in other contexts may be studied.

## Appendix

### A. Frequency Response Optimized Two-Step Adams-Bashforth Methods

The ODE (1) is discretized by a two-step Adams-Bashforth method-like numerical integrator [15] as

$$x_t = a_{-1} x_{t-h} + b_{-1} \dot{x}_{t-h} + b_{-2} \dot{x}_{t-2h} \quad (27)$$

The modified two-step Adams-Bashforth method is expressed as [6]

$$a_{-1} = 1 \quad (28)$$

$$b_{-1} = \frac{\cos(\omega_{select} h) - \cos(2\omega_{select} h)}{\omega_{select} \sin(\omega_{select} h)} \quad (29)$$

$$b_{-2} = -\frac{1}{\omega_{select}} \tan(\frac{\omega_{select} h}{2}) \quad (30)$$

It can be verified that $j\omega_{select}$, $-j\omega_{select}$ and 0 are a single root of the error expression (6) respectively. Therefore this numerical integrator is accurate for variables dominated by the $\omega_{select}$ component and slow variants. The classical two-step Adams-Bashforth method is expressed as [15]

$$a_{-1} = 1, \quad b_{-1} = \frac{3}{2} h, \quad b_{-2} = -\frac{1}{2} h \quad (31)$$

It can be verified that 0 is a triple root of the error expression (6). Therefore this numerical integrator is rather accurate for slow variants.

### B. Frequency Response Optimized Two-Step Backward Differenctiation Formulas

The ODE (1) is discretized by a two-step backward differentiation formula-like numerical integrator [15] as

$$x_t = a_{-1} x_{t-h} + a_{-2} x_{t-2h} + b_0 \dot{x}_t \quad (32)$$

This paper proposes the modified two-step backward differentiation formula as

$$a_{-1} = \frac{4 + 6\cos(\omega_{select} h) + 2\cos(2\omega_{select} h)}{3 + 4\cos(\omega_{select} h) + 2\cos(2\omega_{select} h)} \quad (33)$$

$$a_{-2} = -\frac{1 + 2\cos(\omega_{select} h)}{3 + 4\cos(\omega_{select} h) + 2\cos(2\omega_{select} h)} \quad (34)$$

$$b_0 = \frac{1}{\omega_{select}} \frac{2\sin(\omega_{select} h) + 2\sin(2\omega_{select} h)}{3 + 4\cos(\omega_{select} h) + 2\cos(2\omega_{select} h)} \quad (35)$$

With these coefficients, $j\omega_{select}$, $-j\omega_{select}$ and 0 are made a single root of the error expression (6) respectively. Therefore this numerical integrator is accurate for variables dominated by the $\omega_{select}$ component and slow variants. The classical two-step backward differentiation formula is expressed as [15]

$$a_{-1} = \frac{4}{3}, \quad a_{-2} = -\frac{1}{3}, \quad b_0 = \frac{2}{3} h \quad (36)$$

It can be verified that 0 is a triple root of the error expression (6). Therefore this numerical integrator is rather accurate for slow variants.